\documentclass[multphys,vecphys]{svmult}

\usepackage{makeidx}     
\usepackage{graphicx}    
\usepackage{multicol}    

\makeindex             
\begin{document}

\title*{The Neutral ISM in Nearby Luminous Compact Blue Galaxies}

\author{C. A. Garland,\inst{1} 
D. J. Pisano,\inst{2}
J. P. Williams,\inst{1}
R. Guzm\'an,\inst{3}
\and
F. J. Castander\inst{4}
}

\institute{Institute for Astronomy, University of Hawai'i, 2680 
Woodlawn Drive, Honolulu, HI 96822  USA
\texttt{garland or jpw @ifa.hawaii.edu}
\and CSIRO Australia Telescope National Facility, P. O. Box 76, Epping NSW 1710, Australia
\texttt{Daniel.Pisano@atnf.csiro.au}
\and Department of Astronomy, University of Florida,
216 Bryant Space Science Center, P.O. Box 112055,
Gainesville, FL 32611  USA
\texttt{guzman@astro.ufl.edu}
\and Institut D'Estudis Espacials de Catalunya/CSIC, Edifici Nexus, Gran Capit\`a, 2-4, 08034
Barcelona, Spain
\texttt{fjc@ieec.fcr.es}
}

\authorrunning{C. A. Garland, D. J. Pisano, J. P. Williams et al.}
\maketitle

\section{Introduction}
\label{sec:1}

Luminous Compact Blue Galaxies (LCBGs) are $\sim$L$^\star$, blue,
high surface brightness, high metallicity, vigorously starbursting
galaxies with an underlying older stellar population \cite{guz96, guz98b}.
They include
a variety of morphological types, such as spiral, polar-ring,
interacting/merging and peculiar galaxies.  They have optical diameters
of a few kpc, but are more luminous and more metal rich than the
Blue Compact Dwarf Galaxies widely studied in the nearby Universe,
e.g. \cite{thu81, tay95}.

When Jangren et al. \cite{jan03} compared intermediate redshift LCBGs
with local normal galaxies, they found that they can be isolated
quantitatively on the basis of color, surface brightness, image
concentration and asymmetry, with color and surface brightness giving
the best leverage for separating LCBGs from normal galaxies.  Specifically, 
LCBGs have B$-$V $<$ 0.6, SBe $<$ 21 mag arcsec$^{-2}$,
and M$_B$ $<$ $-$18.5, 
assuming H$_0$ = 70 km s$^{-1}$ Mpc$^{-1}$.  LCBGs are quite common at intermediate 
redshifts, but by z$\sim$0 their
number density has decreased by a factor of ten.  At z$\sim$1, they have a total
star formation rate density equal to that of grand-design spirals at
that time, but today they contribute negligibly to the star formation rate
density of the Universe \cite{mar94}.  Therefore, LCBGs must undergo 
dramatic evolution.  
From studies at intermediate redshift, Koo et al. \cite{koo94}
and Guzm\'an et al. \cite{guz96} suggest that
some LCBGs may be the progenitors of local low-mass dwarf elliptical
galaxies.  Alternatively, Phillips et al. \cite{phi97} 
and Hammer et al. \cite{ham01} suggest that others
may be more massive disks forming from the center outward to become
local L$^\star$ galaxies.

In order to discriminate between the possible evolutionary
scenarios it is essential to measure the dynamical masses of the
galaxies: Are they as massive as implied by their high luminosities?
It is also essential to measure their gas content for future star formation 
in order to constrain the amount of
fading of their stellar populations.  We have undertaken a survey of
local LCBGs in H {\small{I}} and CO to address these questions.  
The H {\small{I}} provides a 
measure of the dynamical mass, while both H {\small{I}} and CO provide measures
of the gas content: H {\small{I}} for long-term star formation and CO for the 
current
burst of star formation.

\section{Observations}
The current sensitivity of telescopes limits us to detecting CO in
LCBGs within
$\sim$70 Mpc.  We used the first data release of the
Sloan Digital Sky Survey to select our sample of nearby LCBGs, using
Jangren et al.'s \cite{jan03} selection criteria.  Out of the 
$\sim$million galaxies in the first data release, only
$\sim$100 are LCBGs, and only 16 are within 70 Mpc.
To these 16 local LCBGs, we added four more from the literature, for a local LCBG
sample of 20 galaxies.
We observed this sample
in 21 cm H {\small{I}} using the Green Bank Telescope at the 
National Radio Astronomy Observatory in Green Bank, West Virginia in
Winter 2002.  The James Clerk Maxwell Telescope on Mauna Kea, Hawai'i
was used to observe our sample in CO(J=2$-1$) in 2002 - 2003.
All sources were detected in H {\small{I}}; 13 were detected
in CO(J=2$-$1).

\section{Results}
\subsection{Dynamical Masses}
We find that local LCBGs span a wide range of dynamical masses,
from 4 $\times$ 10$^9$ 
to 1 $\times$ 10$^{11}$ M$_\odot$ (measured within R$_{25}$).
Figure 1 compares the dynamical masses
of local LCBGs with intermediate redshift LCBGs and local spiral galaxies
of all Hubble types. Many local LCBGs are 
$\sim$ten times less massive than local galaxies of similar luminosities, 
as found for LCBGs at intermediate redshifts \cite{phi97}.
  However, others are as massive as
local galaxies of similar luminosities.  

\subsection{Gas Depletion Time Scales}

We find our 13 LCBGs detected in CO(J=2$-$1) have molecular gas masses
ranging from 5 $\times$ 10$^7$ 
to 2 $\times$ 10$^9$ M$_\odot$ (assuming a Galactic CO-to-H$_2$ conversion
factor of 1.8
$\times$ 10$^{20}$ cm$^{-2}$ K$^{-1}$ km$^{-1}$ s \cite{dam01}).  Note that
these
are most likely underestimates of the molecular gas masses since we are using
CO(J=2$-$1).
The fraction of molecular to atomic gas mass is small, ranging from 0.03 to 0.3,
similar to local late-type spiral galaxies \cite{rob94}.

We estimated star formation rates from available IRAS data, using
60 and 100 $\mu$m fluxes as outlined in Kewley et al. \cite{kew02}.
The star formation rates for local LCBGs range from $\sim$1 to 
15 M$_\odot$ year$^{-1}$.  For comparison, local spirals of all
types have star formation rates $\sim$2 M$_\odot$ year$^{-1}$ 
\cite{rob94}, so these
LCBGs do not have unusually high star formation rates.  However, they do
have very high \emph{specific} star formation rates--the ratio of star
formation rate to dynamical mass (within R$_e$).  As seen in Figure 2, local LCBGs
have specific star formation rates from $\sim$3 to 40 times those of local normal
spirals \cite{rob94}.  The specific star formation rates of local LCBGs are in 
the same range as local H~{\small{II}} (starbursting) galaxies \cite{guz97}.

We find that the molecular gas in local LCBGs
is depleted quite quickly, in 30 to 200 million years.
The 
molecular plus atomic gas is depleted in 30 million to 10 billion years;
however, $\sim$80\% of the local LCBGs deplete their gas in less than 5 billion years.
Therefore, most LCBGs will not be able to sustain their current rates of 
star formation and will eventually fade.  

\section{Conclusions}
Both in dynamical masses and gas depletion time scales, we find that
local LCBGs have a wide range of characteristics and are unlikely to
evolve into one galaxy class. They have dynamical masses consistent with
a range of galaxy types, such as dwarf ellipticals, Magellanic (low-luminosity)
spirals and normal spirals.  The majority have atomic plus molecular gas 
depletion time scales less than five billion years;  such galaxies
may have masses, sizes and faded luminosities and surface brightnesses
consistent with the brightest local dwarf ellipticals.  A few local LCBGs
have longer gas depletion time scales, approaching a Hubble time.  These
may fade very little, becoming spirals or Magellanic irregulars.

\smallskip
\noindent
\emph{Acknowledgements}
Support for this work was provided by the NSF
through award GSSP02-0001 from the NRAO.
Support for conference
attendance was provided by the AAS and NSF in the form of an International Travel Grant. 

\begin{figure}
\centering
\includegraphics[height=7cm]{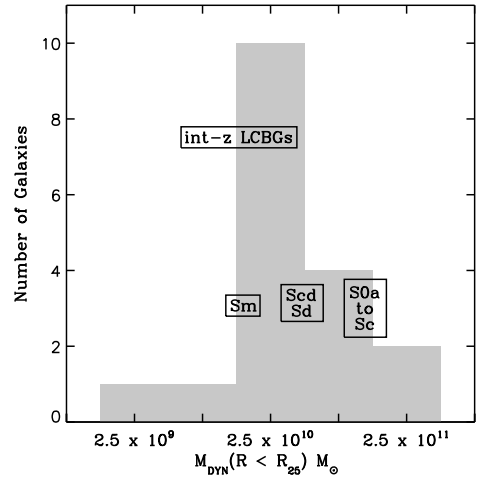}
\caption{Dynamical masses (within R$_{25}$) for local
LCBGs, as measured from 
H~{\small{I}} observations.
For comparison, the ranges of dynamical masses for 
intermediate redshift LCBGs \cite{phi97}, 
and local spiral galaxies \cite{rob94} are
indicated.  Note that ``Sm'' indicates Magellanic
or low-luminosity spirals.}
\label{fig:1}       
\end{figure}

\begin{figure}
\centering
\includegraphics[height=7cm]{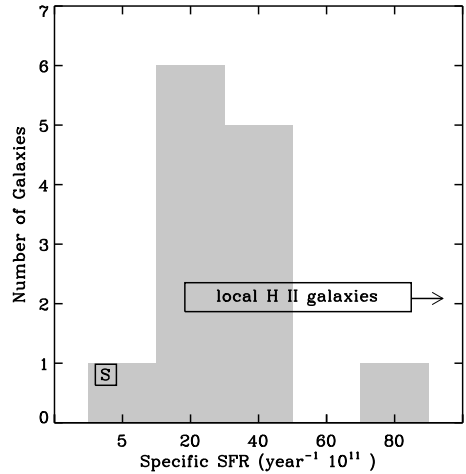}
\caption{The \emph{specific} star formation rate 
(ratio of star formation rate to dynamical mass within Re) for
the local sample of LCBGs.  Their specific star formation rates
are much higher than all types of local spirals
(S); they are similar to local H {\small{II}} galaxies.}
\label{fig2}       
\end{figure}


\printindex
\end{document}